\documentclass[12pt]{article}
\pdfoutput=1

\usepackage{jheppub}
\usepackage{amsmath,amssymb,euscript,array,mathrsfs}
\usepackage{slashed}

\usepackage{epsfig}

\usepackage{tikz}
\usetikzlibrary{matrix,arrows,positioning,shapes,snakes,fadings,decorations.pathmorphing,
decorations.pathreplacing,automata,shadows,decorations.markings,patterns}
\usetikzlibrary{calc,shapes.callouts,shapes.arrows}

\def\a{\alpha}
\def\b{\beta}
\def\c{\gamma}
\def\d{\delta}
\def\e{\epsilon}

\def\l{\lambda}
\def\m{\mu}
\def\n{\nu}

\def\r{\rho}
\def\s{\sigma}
\def\t{\tau}

\def\w{\omega}

\def\D{\Delta}

\def\S{\Sigma}

\def\tr{{\rm tr}}

\def\Dbarslash{\,\,{\raise.15ex\hbox{/}\mkern-12mu {\bar D}}}
\def\Dslash{\,\,{\raise.15ex\hbox{/}\mkern-12mu D}}
\def\delslash{\,\,{\raise.15ex\hbox{/}\mkern-9mu \partial}}
\def\delbarslash{\,\,{\raise.15ex\hbox{/}\mkern-9mu {\bar\partial}}}

\def\tr{{\rm tr}}

\def\Im{{\rm Im}}

\title{\begin{center}A Model of Gravitational Leptogenesis\end{center}}

\author{Graham M. Shore}

\emailAdd{g.m.shore@swansea.ac.uk}

\affiliation{Department of Physics, College of Science, Swansea
  University, Singleton Park, Swansea, SA2 8PP, UK.}

\date{today}

\abstract{Gravitational leptogenesis is an elegant way of explaining the matter-antimatter
asymmetry in the universe. This paper is a review of the recently proposed mechanism
of radiatively-induced gravitational leptogenesis (RIGL), in which loop effects in QFT in 
curved spacetime automatically generate an asymmetry between leptons and antileptons
in thermal quasi-equilibrium in the early universe. The mechanism is illustrated in a
simple see-saw BSM model of neutrinos, where the lepton-number violating interactions
required by the Sakharov conditions are mediated by right-handed neutrinos with 
Majorana masses of $O(10^{10})\,{\rm GeV}$. The Boltzmann equations are extended to include
new, loop-induced gravitational effects and solved to describe the evolution of the 
lepton number asymmetry in the early universe. With natural choices of neutrino parameters,
the RIGL mechanism is able to generate the observed baryon-to-photon ratio 
in the universe today. 

\vskip4.5cm }

\notoc

\begin{document}

\maketitle
\flushbottom

\setlength{\parskip}{10pt}

\section{Radiatively-Induced Gravitational Leptogenesis -- overview}\label{sect 1}

The origin of the matter-antimatter asymmetry of the universe is one of the most 
important outstanding issues in cosmology. In this
review,\footnote{This paper is an edited arXiv version of a contribution
to the Festschrift celebrating the physics career of Peter Suranyi.}  based on the recent
paper \cite{McDonald:2020ghc}, we describe a proposal
in which a lepton number asymmetry is generated dynamically in a minimal extension
of the standard model by gravitational effects in the expanding universe. 

The fundamental mechanism, {\it radiatively-induced gravitational leptogenesis}, 
was introduced in the series of papers \cite{McDonald:2014yfg, McDonald:2015ooa, 
McDonald:2015iwt,McDonald:2016ehm, McDonald:2020ghc}. The key idea is that in a minimal 
extension of the standard model to include heavy right-handed neutrinos $\nu_R^{\,\a}$
($\a = 1,2,3$), 
an asymmetry in the propagation of the light neutrinos $\nu_L^{\,i}$ ($i = e,\m,\t$)
and antineutrinos arises due to gravitational tidal effects in loop diagrams 
in which the heavy neutrinos appear as virtual intermediate states. This induces 
an effective gravitational interaction in which the rate of change of the Ricci scalar
plays the role of a chemical potential for lepton number \cite{Davoudiasl:2004gf}. 
In the high temperature environment of the early universe, this generates a 
lepton number asymmetry in quasi-equilibrium. Later, as the universe expands and cools, 
this lepton asymmetry is transformed into the observed baryon asymmetry through 
the well-known sphaleron mechanism \cite{Khlebnikov:1988sr}.

These loop effects violate the strong equivalence principle (SEP) of classical general 
relativity, and are most usefully expressed in terms of a weak-curvature effective
action which includes SEP-violating direct couplings of the curvature tensor to the
light lepton fields. The curvature couplings involve inverse powers of the heavy neutrino
masses $M_\a$, since these set the effective length scale of the virtual loops on
which the tidal gravitational forces act.  

The necessary conditions for a successful model of baryogenesis, or leptogenesis, were
set out long ago in the famous Sakharov conditions \cite{Sakharov:1967dj}, according to 
which the fundamental theory should exhibit (i) baryon or lepton violation, (ii) C and CP
violation, and (iii) the mechanism must involve {\it non-equilibrium} dynamics.

The fundamental BSM theory we consider here is the familiar extension of the standard
model with sterile right-handed neutrinos ({\it i.e.}~coupling only to the 
Higgs field and the light leptons through Yukawa interactions, and neutral
under the gauge interactions of the SM). We include 
Majorana mass terms of $O(10^{10})\,{\rm GeV}$ and above for the right-handed 
neutrinos, which allows the observed light neutrino masses to be generated by the
see-saw mechanism. The addition of the right-handed neutrinos is the only BSM
ingredient - our leptogenesis mechanism then follows automatically and necessarily 
from the gravitational dynamics at one and two-loop order in this model.  

The BSM Lagrangian is therefore
\begin{equation}
S\,=\, \int d^4 x \sqrt{-g} \biggl[{\cal L}_{\rm SM} + \biggl(
\frac{1}{2} i\,\overline{\n_R}\, \c.\overleftrightarrow{D} \n_R - \frac{1}{2} 
\overline{\n_R^{\,\,c}}\, M\,\n_R
- \overline{\ell_L} \,\l \,\phi\, \n_R + {\rm h.c.} \biggr)\biggr] \ .
\label{a1}
\end{equation}
Here, $M_{\a\b}$ is the Majorana mass matrix, which we take to be diagonal. 
The fields $\ell_L^i$ \,($i = e, \m, \t$) are the SM lepton doublets
and $\phi$ is the Higgs field.\footnote{The notation is $\phi_r = \e_{rs}\tilde\phi_s^*$ 
where $\tilde\phi$ is the usual Higgs doublet giving mass to the lower fields in the 
$SU(2)$ lepton doublets.} The Yukawa couplings $\l_{i\a}$ are complex and are
responsible for CP violation in this model. These flavour indices are suppressed 
here and in (\ref{a2}), (\ref{a3}) below.

The inclusion of Majorana masses for the sterile neutrinos allows the lepton number 
violating scattering processes $\n_L\, H \leftrightarrow \n_L^{\,c}\, H$ and 
$\n_L\, \n_L \leftrightarrow H\, H$,
where $H$ is the physical Higgs boson, as required by the first Sakharov condition.
As is well-known, the model also exhibits a non-gravitational mechanism for generating
a lepton asymmetry -- thermal leptogenesis-- through the {\it out-of-equilibrium}
lepton number violating decays of the heavy neutrinos 
$\n_R \rightarrow \n_L H$ \cite{Fukugita:1986hr}.
See, for example, ref.~\cite{Buchmuller:2004nz} for a review.

In the radiatively-induced gravitational leptogenesis (RIGL) mechanism, the 
third Sakharov condition is replaced by the time-dependence of the background 
gravitational field. Tidal effects, arising first at two-loop order in this model,
generate the following CP odd effective interaction of the lepton (neutrino) 
number current $J^\m = \overline{\n_L} \c^\m \n_L$ with the rate of change
of the Ricci scalar,
\begin{equation}
S_b \,=\, \int d^4 x \sqrt{-g}\,\, b\, \partial_\m R\, J^\m \ ,
\label{a2}
\end{equation}
where the coupling $b$ is $O(\l^4/M^2)$.
Evidently, a time-dependence of the Ricci scalar reduces this interaction to
$b\, \dot{R}\, Q$, where $Q$ is the lepton number charge, which is equivalent
to adding a chemical potential $\m = b\, \dot{R}\,$.\footnote{Note that in the original 
gravitational baryogenesis (leptogenesis) proposal of 
Davoudiasl {\it et al.}~\cite{Davoudiasl:2004gf}, 
the interaction (\ref{a2}) with $J^\m$ interpreted as the baryon (lepton) number current
is postulated essentially {\it a priori}, with an arbitrary mass-dependent coupling. 
Here, we show how it is necessarily generated in the effective action of the simple 
and phenomenologically motivated BSM model with heavy Majorana masses for the right-handed 
neutrinos. In an earlier proposal of `spontaneous baryogenesis'
\cite{Cohen:1987vi}, an interaction 
of the form (\ref{a2}) but with the Ricci scalar replaced by a time-dependent background
scalar field was proposed. } 
At finite temperature, this will induce a non-vanishing lepton number density
$n_L^{eq} = \tfrac{1}{3}b\, \dot{R}\, T^2$ in quasi-equilibrium.

Further, CP even, interactions are generated already at one-loop level and the complete 
effective action for the light neutrinos is
\begin{align}
S_{eff}= \int d^4 x\sqrt{-g} \biggl[ 
\frac{1}{2} i\,\overline{\n_L} \c.\overleftrightarrow{D}\, \n_L
&\,+ \,(a - \tfrac{1}{2} d) \, R_{\m\n} \, i \,\overline{\n_L} \c^\m 
\overleftrightarrow{D}^\n \,\n_L
\,+\, b \,\partial_\m R \,\,\overline{\n_L} \c^\m \,\n_L  \nonumber \\
&~~~\,+ c \, R \, i \,\overline{\n_L} \c.\overleftrightarrow{D} \,\n_L
\,-\, d\, i\left(D_\m \overline{\n_L}\right) \c.\overleftrightarrow{D} \, 
D^\m \n_L \biggr] \ .
\label{a3}
\end{align}
Here, $a_{ij}, c_{ij}, d_{ij}$ are coefficients of $O(\l^2/M^2)$, and we have suppressed
the corresponding $i,j$ flavour indices in the fermion bilinears for simplicity of notation.
The effective action is valid to lowest order in $\mathfrak{R}/M^2$, where 
$\mathfrak{R}$ denotes a typical curvature component, and also in the `low-energy'
regime $E\sqrt{\mathfrak{R}}/M^2 \ll 1$  (see ref.~\cite{McDonald:2020ghc} for a careful
discussion, including the justification for using the effective Lagrangian at temperatures
in excess of $M$).

These CP even interactions modify the conservation law of the lepton number current,
giving
\begin{equation}
D_\m J^\m \,=\, - 2a \, R_{\m\n} D^\m J^\n \, - \, 2 \hat{b} \, \partial_\m R J^\m \ ,
\label{a4}
\end{equation}
where $\hat{b} = \tfrac{1}{2}a + c + \tfrac{1}{4}d$.
In a FRW universe, this implies the following equation for the time evolution 
of the lepton number density $n_L$, 
\begin{equation}
\frac{dn_L}{dt} \,+\, 3H n_L \,+\, 2a \left(-3 R^0{}_0 + R^i{}_i\right) H n_L \,+\, 
2 \hat{b} \,\dot{R} \,n_L \,=\, 0 \ ,
\label{a5}
\end{equation}
where $H$ is the Hubble parameter. 
This shows that the lepton number density $n_L$ evolves non-trivially in a gravitational
background due to quantum loop effects.
Whether this effect tends to amplify or suppress the magnitude of $n_L$ depends on the signs
and relative magnitudes of the coefficients, especially $a$, which are not arbitrary in RIGL
but are determined by the fundamental BSM theory.

Together with the gravitationally-induced equilibrium lepton asymmetry $n_L^{eq}$,
these two effects combine to generate a novel dynamical evolution \cite{McDonald:2020ghc} of 
the lepton number density in the expanding universe. To describe this quantitatively in the
high temperature environment of the early FRW universe, we need the corresponding
Boltzmann equations. Neglecting for the moment the contribution to the asymmetry from
decays of the heavy $\n_R$ neutrinos, the gravitationally modified Boltzmann equation is
\begin{equation}
\frac{dN_L}{dz} \,=\, -\, W \, \bigl(N_L\,-\, N_L^{eq}\bigr) \,-\, \mathcal{W}\, N_L \ .
\label{a6}
\end{equation}
Here, $N_L(z) = n_L/n_\c$ is the ratio of the number density of light leptons to photons, 
and we have replaced the time variable with $z = M_1/T$, where $T$ is temperature
and $M_1$ is the mass of the dominant heavy neutrino. 

The new gravitational terms in (\ref{a6}) are the equilibrium number asymmetry $N_L^{eq}(z)$
generated by the interaction (\ref{a2}) and the evolution term $\mathcal{W}(z)$ arising from
the CP even interactions in (\ref{a3}), which reflects the curvature-induced evolution terms 
in (\ref{a5}). $W(z) = \Gamma/zH$ is the usual factor
determined by the interaction rate $\Gamma(z)$ for the lepton number violating interactions.
Without the new gravitational terms, it acts as a `washout' factor -- here, it has the
different role of driving the lepton number asymmetry towards its {\it non-vanishing}
equilibrium value $N_L^{eq}(z)$.

\begin{figure}[h!]
\vskip0.5cm
\centering{\includegraphics[scale=1.2]{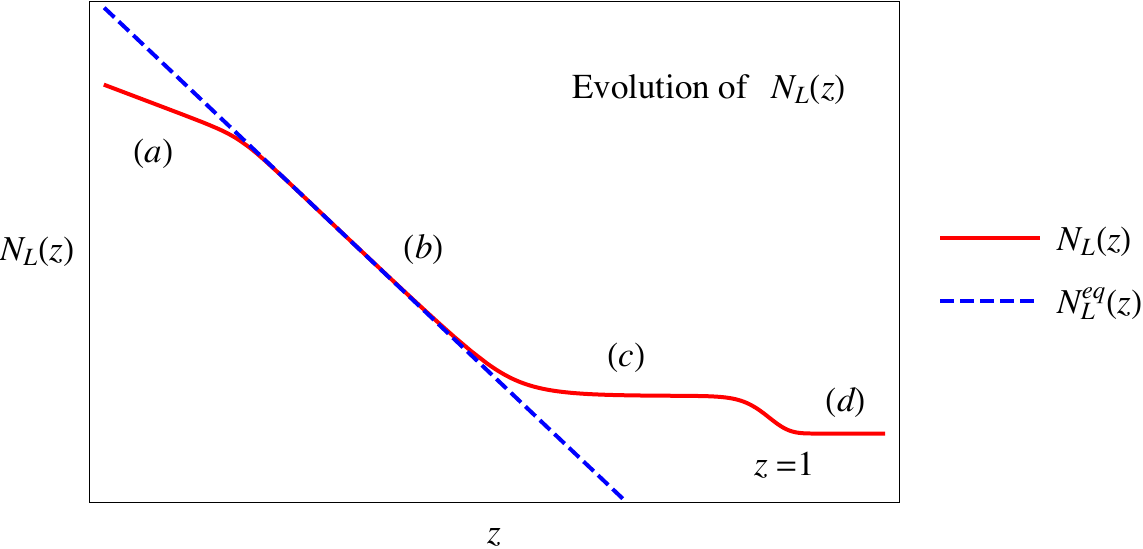}}
\caption{Illustration of the key stages in the evolution of the lepton asymmetry $N_L(z)$
with temperature, given by the Boltzmann equation (\ref{a6}). The temperature dependence of
the normalised rate factor $W(z)$ for the lepton-violating reactions, which drive the asymmetry
towards its equilibrium value $N_L^{eq}(z)$, is $W(z) \sim 1/z^2$, while both the 
gravitationally-induced terms $N_L^{eq}(z)$ and $\mathcal{W}(z)$ fall off as $1/z^5$. }
\label{Fig a.1}
\end{figure}

Putting all this together, the entire evolution predicted by the RIGL Boltzmann equation is 
shown in Fig.~\ref{Fig a.1}. Essentially we find four stages. First, there is a very early high 
temperature phase (region (a)) in which the new evolution term $\mathcal{W}(z)$ keeps $N_L$ 
below equilibrium, followed by a phase where $W(z) > \mathcal{W}(z)$ and $N_L$ is driven to 
its gravitationally-induced equilibrium value $N_L^{eq}(z)$ (region (b)). Next is the vital 
decoupling transition (region (c)) when $W(z)$ becomes too weak to hold $N_L$ to $N_L^{eq}(z)$ 
({\it i.e.}~the $L$-violating reactions are too slow compared 
to the Hubble expansion to maintain equilibrium) and it decouples leaving a constant, frozen  
asymmetry. Finally, there is a late dip in $N_L$ as 
$W(z)$ develops a resonance peak around $T\sim M_1$, temporarily pushing $N_L$ lower 
towards the rapidly decreasing $N_L^{eq}(z)$ (region (d)).

The transitions between these stages depend on the dynamical balance between 
the temperature and curvature dependent rates $W(z)$  and $\mathcal{W}(z)$, and 
$N_L^{eq}(z)$, as the universe expands and cools. In particular, the sign of $\mathcal{W}(z)$
is key to whether the new evolution term amplifies or suppresses the lepton asymmetry 
at early times. In the BSM model described here, the sign is such as to suppress the
asymmetry, as shown in Fig.~\ref{Fig a.1}. 

In the remainder of this paper, we describe this scenario in more detail and describe
the quantitative predictions for the baryon asymmetry of the universe in this model.

\section{Matter-Antimatter Asymmetry from Gravitational Interactions}\label{sect 2}

In this section, we explain in principle how gravitational interactions can induce an 
asymmetry in the dynamics of matter and antimatter, and outline the methods used to 
obtain the gravitational effective action (\ref{a3}) from the fundamental BSM theory
(\ref{a1}).

It is useful to begin by rewriting the action (\ref{a1}) explicitly in terms of the
fields $\n_L$ and $\n_L^{\,c}$ corresponding to the light neutrinos and their 
antiparticles\footnote{The charge conjugate field is defined as 
$\n_L^{\,c} \equiv (\n_L)^c = -i \c^2 \n_L^{\,*}$ and is right-handed. With a Higgs VEV $v$,
the see-saw mechanism gives rise to three light Majorana neutrinos with mass matrix
\begin{equation*}
(m_\n)_{ij} ~=~ \sum_\a \l_{i\a}\,\frac{1}{M_\a}\,\l_{\a j}^T\, v^2 \ ,
\end{equation*}
along with the heavy sterile Majorana neutrinos with masses $M_\a$. This is diagonalised by the 
PMNS matrix $U$ such that $U^T m_\n U = m_{diag}$, with the corresponding relation between the
mass and flavour eigenstates.}
\footnote{The coupling to gravity is through the connection alone, 
which is the requirement for the classical Lagrangian to satisfy the strong
equivalence principle. The covariant derivative acting on spinors is 
$D_\m = \partial_\m - \tfrac{i}{4}\w_{\m a b}\s^{a b}$, where $\w_{\m a b}$ is the spin connection
and $\s^{ab} = \tfrac{i}{2} \left[\c^a,\c^b\right]$.}
\begin{align}
S\,=\, \int d^4 x \sqrt{-g}&\biggl[{\cal L}_{\rm SM} + 
\frac{1}{4}\left(i\,\overline{\n_R} \,\c.\overleftrightarrow{D}\, \n_R  
+ i\, \overline{\n_R^{\,\,c}}\,\c.\overleftrightarrow{D}\,\n_R^{\,\,c}\right) 
- \frac{1}{2} \Bigl(\overline{\n_R^{\,\,c}}\, M\, \n_R  +
\overline{\n_R} \,M \,\n_R^{\,\,c} \Bigr) 
\nonumber \\
&-\frac{1}{2} \left(\overline{\ell_L}\, \l \,\phi \,\n_R + \overline{\n_R}\, 
\l^\dagger \phi^\dagger\, \ell_L
+ \overline{\ell_L^{\,\,c}} \,\l^* \phi^*\, \n_R^{\,\,c} + \overline{\n_R^{\,\,c}}
 \,\l^T \phi^T \,\ell_L^{\,\,c} \right) \biggr] \ .
\label{b1}
\end{align}

\begin{figure}[h!]
\begin{center}
\begin{tikzpicture}[scale=0.74]
\draw[thick] (0,-1.5) -- (1.5,0);
\draw[thick,dash pattern = on 6pt off 4pt ] (0,1.5) -- (1.5,0);
\draw[thick] (1.5,0) -- (4.5,0);
\draw[thick] (4.5,0) -- (6.0,-1.5);
\draw[thick,dash pattern = on 6pt off 4pt ] (4.5,0) -- (5.95,1.5);
\filldraw[black] (1.5,0) circle (1.5pt);
\filldraw[black] (4.5,0) circle (1.5pt);
\node at (0.0,-1.0) { \small $\nu_L$}; 
\node at (0.5,1.6) { \small $H$};
\node at (6.1,-1.0) { \small $\nu_L^{\,c}$};
\node at (5.5,1.6) { \small $H$};
\node at (3.0,0.4) { \small $S_\alpha^\times$};
\end{tikzpicture} 
\hskip1.4cm
\begin{tikzpicture}[scale=0.74]
\draw[thick] (0,1.0) -- (2.5,1.0);
\draw[thick,dash pattern = on 6pt off 4pt ] (2.5,1.0) -- (4.95,1.0);
\draw[thick] (2.5,1.0) -- (2.5,-1.5);
\draw[thick] (2.5,-1.5) -- (5,-1.5);
\draw[thick,dash pattern = on 6pt off 4pt ] (0,-1.5) -- (2.5,-1.5);
\filldraw[black] (2.5,1.0) circle (1.5pt);
\filldraw[black] (2.5,-1.5) circle (1.5pt);
\node at (0.2,1.4) { \small $\nu_L$}; 
\node at (4.8,1.4) { \small $H$};
\node at (4.8,-1.1) { \small $\nu_L^{\,c}$};
\node at (0.2,-1.1) { \small $H$};
\node at (2.9,-0.25) { \small $S_\alpha^\times$};
\end{tikzpicture} 
\vskip1.4cm
\begin{tikzpicture}[scale=0.74]
\draw[thick] (0,1.25) -- (2.5,1.25);
\draw[thick,dash pattern = on 6pt off 4pt ] (2.5,1.25) -- (4.95,1.25);
\draw[thick] (2.5,1.25) -- (2.5,-1.25);
\draw[thick] (0,-1.25) -- (2.5,-1.25);
\draw[thick,dash pattern = on 6pt off 4pt ] (2.5,-1.25) -- (4.95,-1.25);
\filldraw[black] (2.5,1.25) circle (1.5pt);
\filldraw[black] (2.5,-1.25) circle (1.5pt);
\node at (0.2,1.65) { \small $\nu_L$}; 
\node at (4.8,1.65) { \small $H$};
\node at (4.8,-0.9) { \small $H$};
\node at (0.2,-0.9) { \small $\nu_L$};
\node at (2.9,0) { \small $S_\alpha^\times$};
\end{tikzpicture} 
\hskip2.75cm
\begin{tikzpicture}[scale=0.74]
\draw[thick] (0,1.25) -- (2.5,1.25);
\draw[thick,dash pattern = on 6pt off 4pt ] (2.5,1.25) -- (5.15,-1.25);
\draw[thick] (2.5,1.25) -- (2.5,-1.25);
\draw[thick] (0,-1.25) -- (2.5,-1.25);
\draw[thick,dash pattern = on 6pt off 4pt ] (2.5,-1.25) -- (4.97,1.25);
\filldraw[black] (2.5,1.25) circle (1.5pt);
\filldraw[black] (2.5,-1.25) circle (1.5pt);
\node at (0.2,1.65) { \small $\nu_L$}; 
\node at (5.15,1.55) { \small $H$};
\node at (5.4,-0.9) { \small $H$};
\node at (0.2,-0.9) { \small $\nu_L$};
\node at (2.9,0) { \small $S_\alpha^\times$};
\end{tikzpicture} 
\end{center}
\caption{Feynman diagrams for the $\D L=2$ lepton number violating reactions 
$\n_L~H \leftrightarrow \n_L^{\,c}~ H$ and $\n_L~\n_L \leftrightarrow H ~H$, mediated by 
the `charge violating' $\langle\, \n_R ~~ \overline{\n_R^{\,\,c}}\,\rangle$ propagator 
$S_\a^\times$. }
\label{Fig b.1}
\end{figure}
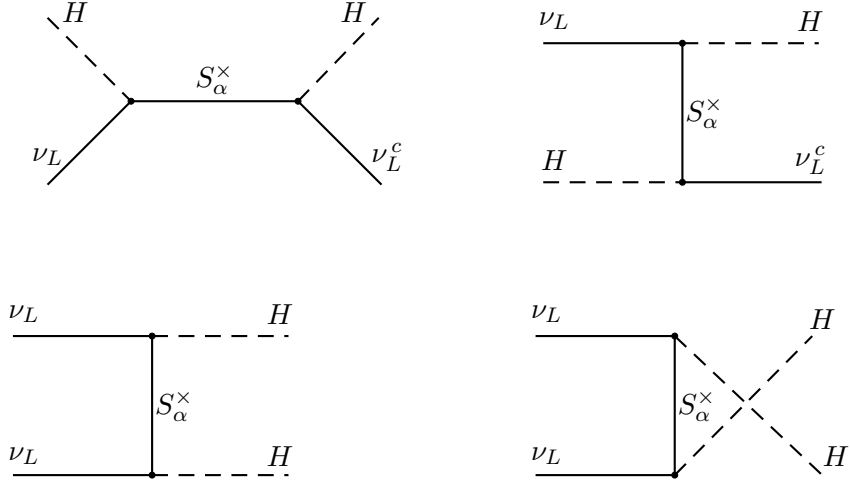

The Majorana mass for the right-handed neutrinos allows the occurrence of both
leptonic ``{\it charge-conserving}'' and ``{\it charge-violating}'' propagators:
\begin{align}
\langle\, \n_R ~~ \overline{\n_R}\,\rangle ~&=~ \langle\,\n_R^{\,\,c} ~~ 
\overline{\n_R^{\,\,c}}\,\rangle 
~=~ S(x,y) \ , \\
\langle\, \n_R ~~ \overline{\n_R^{\,\,c}}\,\rangle ~&=~ \langle\,\n_R^{\,\,c} ~~ 
\overline{\n_R}\,\rangle 
~=~ S^\times(x,y) \ .
\label{b2}
\end{align}
In flat spacetime, these are
\begin{equation}
S_\a(p) ~=~ \frac{i\,\c.p}{p^2 - M_\a^2} \ , ~~~~~~~~~~~~~~~~~ 
S_\a^\times(p) ~=~ \frac{i\, M_\a}{p^2 - M_\a^2} \ .
\label{b3}
\end{equation}

The existence of the charge-violating $S^\times(x,y)$ propagator plays a key role in 
generating the lepton asymmetry. First, it allows the $\D L =2$ scattering reactions 
$\n_L~H \leftrightarrow \n_L^{\,\,c}~ H$ and 
$\n_L~\n_L \leftrightarrow H ~H$, illustrated in Fig.~\ref{Fig b.1}. These diagrams depend on the 
Yukawa coupling factor through $\l \,S^\times \l^T \,=\, \sum_\a \l_{i\a} S_\a^\times \l_{\a j}^T$.
This is the source of the lepton number violation required by the first Sakharov condition.

Next, to implement the RIGL mechanism, we need to show that the propagation of leptons 
and antileptons is different in a gravitational field.
Specifically, we find that at two-loop level, the self-energies $\S$ and $\S^c$ for the leptons and
antileptons differ when translation invariance no longer holds, leading to distinct dispersion
relations. 

\begin{figure}[h!]
\begin{center}
\begin{tikzpicture}[scale=1]
\draw[thick] (0,0) -- (2,0);
\draw[thick] (2,0) -- (5,0);
\draw[thick] (5,0) -- (7,0);
\draw[dash pattern = on 6pt off 4pt,thick] (5,0) arc (15:165:1.55); 
\filldraw[black] (2,0) circle (1.5pt);
\filldraw[black] (5,0) circle (1.5pt);
\node at (0.2,-0.4) { \small $\ell_L$}; 
\node at (1.6,-0.32) { \small $\overline{\ell_L}$}; 
\node at (2.4,-0.4) { \small $\nu_R$}; 
\node at (4.6,-0.4) { \small $\overline{\nu_R}$}; 
\node at (5.4,-0.4) { \small $\ell_L$}; 
\node at (6.8,-0.32) { \small $\overline{\ell_L}$}; 
\node at (3.4,-0.45) { \small $S_\alpha$}; 
\node at (1.8,0.5) { \small $\phi$}; 
\node at (5.2,0.5) { \small $\phi^\dagger$}; 
\end{tikzpicture}  
\end{center}
\caption{ One-loop self-energy diagram for the light $\n_L$ neutrinos with an intermediate
charge-conserving $\n_R$ propagator $S$.  The fields at the vertices are shown explicitly for 
comparison with the Lagrangian (\ref{b1}).}
\label{Fig b.2}
\end{figure}
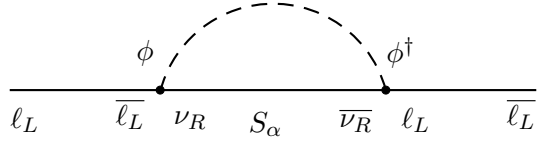

At one loop, there is a single self-energy diagram for the light neutrinos $\n_L$ 
(Fig.~\ref{Fig b.2}), which involves the charge-conserving right-handed neutrino 
propagator $S_\a$. 
(For neutrino self-energies in the
standard model, see refs.~\cite{Ohkuwa:1980jx, McDonald:2014yfg}) 
There is no corresponding one-loop diagram with the charge-violating
propagator $S_\a^\times$. The self-energy is therefore,
\begin{equation}
\S_{ij}(x,y) ~=~ \sum_\a \l_{i\a} \l_{aj}^\dagger\, G(x,y)\, S_\a (x,y) \ .
\label{b4}
\end{equation}
A similar diagram gives the self-energy for the antineutrinos. 
Since we are interested in the violation of total lepton number, we trace over the light lepton
flavours, and find
\begin{align}
\tr \left(\S_{ij}(x,y) - \S_{ij}^c(x,y)\right) ~&=~ 
\sum_\a \left(\l^\dagger \l - \l^T \l^*\right)_{\a\a} \, G(x,y)\, S_\a (x,y)  \nonumber \\
&=~ 2i\, \sum_\a \Im \left(\l^\dagger \l\right)_{\a\a} \,  G(x,y)\, S_\a (x,y) ~=~ 0 \ .
\label{b5}
\end{align}

These diagrams therefore imply that the gravitational influence on the propagation of the light 
neutrinos and antineutrinos is identical. In the effective Lagrangian (\ref{a3}), they correspond 
to the CP even terms. However, they do not contribute to the CP odd term which is responsible
for generating the lepton-antilepton asymmetry. For this, we require the two-loop self-energies,
which have the $O(\l^4)$ dependence necessary to exhibit CP violation.

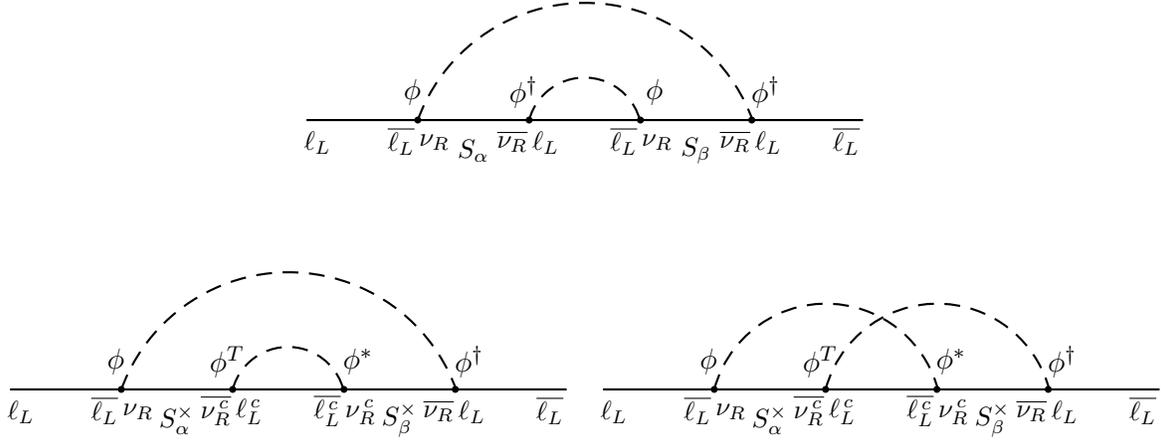
\begin{figure}[h!]
\begin{center}
\begin{tikzpicture}[scale=0.74,decoration={markings,mark=at position 0.52 with {\arrow[scale=1.2,>=latex]{>}}}]
\draw[thick] (0,0) -- (2,0);
\draw[thick] (2,0) -- (4,0);
\draw[thick] (4,0) -- (6,0);
\draw[thick] (6,0) -- (8,0);
\draw[thick] (8,0) -- (10,0);
\draw[dash pattern = on 6pt off 4pt,thick] (2,0) arc (160:20:3.20);
\draw[dash pattern = on 6pt off 4pt,thick] (4,0) arc (165:15:1.03);
\filldraw[black] (2,0) circle (1.5pt);
\filldraw[black] (4,0) circle (1.5pt);
\filldraw[black] (6,0) circle (1.5pt);
\filldraw[black] (8,0) circle (1.5pt);
\node at (0.2,-0.4) { \footnotesize $\ell_L$}; 
\node at (1.7,-0.38) { \footnotesize $\overline{\ell_L}$}; 
\node at (2.3,-0.4) { \footnotesize $\nu_R$}; 
\node at (3.7,-0.38) { \footnotesize $\overline{\nu_R}$}; 
\node at (4.3,-0.4) { \footnotesize $\ell_L$}; 
\node at (5.7,-0.38) { \footnotesize $\overline{\ell_L}$}; 
\node at (6.3,-0.4) { \footnotesize $\nu_R$}; 
\node at (7.7,-0.38) { \footnotesize $\overline{\nu_R}$}; 
\node at (8.3,-0.4) { \footnotesize $\ell_L$}; 
\node at (9.7,-0.38) { \footnotesize $\overline{\ell_L}$}; 
\node at (3.0,-0.55) { \footnotesize $S_\alpha$}; 
\node at (7.0,-0.55) { \footnotesize $S_\beta$};
\node at (1.9,0.5) { \small $\phi$};
\node at (3.9,0.5) { \small $\phi^\dagger$};
\node at (6.25,0.5) { \small $\phi$};
\node at (8.25,0.5) { \small $\phi^\dagger$};
\end{tikzpicture}  
\end{center}
\vskip0.5cm  
\begin{tikzpicture}[scale=0.74,decoration={markings,mark=at position 0.52 with {\arrow[scale=1.2,>=latex]{>}}}]
\draw[thick] (0,0) -- (2,0);
\draw[thick] (2,0) -- (4,0);
\draw[thick] (4,0) -- (6,0);
\draw[thick] (6,0) -- (8,0);
\draw[thick] (8,0) -- (10,0);
\draw[dash pattern = on 6pt off 4pt,thick] (2,0) arc (160:20:3.20);
\draw[dash pattern = on 6pt off 4pt,thick] (4,0) arc (165:15:1.03);
\filldraw[black] (2,0) circle (1.5pt);
\filldraw[black] (4,0) circle (1.5pt);
\filldraw[black] (6,0) circle (1.5pt);
\filldraw[black] (8,0) circle (1.5pt);
\node at (0.2,-0.4) { \footnotesize $\ell_L$}; 
\node at (1.7,-0.38) { \footnotesize $\overline{\ell_L}$}; 
\node at (2.3,-0.4) { \footnotesize $\nu_R$}; 
\node at (3.7,-0.38) { \footnotesize $\overline{\nu_R^{\,c}}$}; 
\node at (4.3,-0.4) { \footnotesize $\ell_L^{\,c}$}; 
\node at (5.7,-0.38) { \footnotesize $\overline{\ell_L^{\,c}}$}; 
\node at (6.3,-0.4) { \footnotesize $\nu_R^{\,c}$}; 
\node at (7.7,-0.38) { \footnotesize $\overline{\nu_R}$}; 
\node at (8.3,-0.4) { \footnotesize $\ell_L$}; 
\node at (9.7,-0.38) { \footnotesize $\overline{\ell_L}$}; 
\node at (3.0,-0.55) { \footnotesize $S_\alpha^\times$}; 
\node at (7.0,-0.55) { \footnotesize $S_\beta^\times$};
\node at (1.9,0.5) { \small $\phi$};
\node at (3.9,0.5) { \small $\phi^T$};
\node at (6.25,0.5) { \small $\phi^*$};
\node at (8.25,0.5) { \small $\phi^\dagger$};
\end{tikzpicture}  
\hskip0.1cm
\begin{tikzpicture}[scale=0.74,decoration={markings,mark=at position 0.52 with {\arrow[scale=1.2,>=latex]{>}}}]
\draw[thick] (0,0) -- (2,0);
\draw[thick] (2,0) -- (4,0);
\draw[thick] (4,0) -- (6,0);
\draw[thick] (6,0) -- (8,0);
\draw[thick] (8,0) -- (10,0);
\draw[dash pattern = on 6pt off 4pt,thick] (2,0) arc (165:15:2.08);
\draw[dash pattern = on 6pt off 4pt,thick] (4,0) arc (165:15:2.08);
\filldraw[black] (2,0) circle (1.5pt);
\filldraw[black] (4,0) circle (1.5pt);
\filldraw[black] (6,0) circle (1.5pt);
\filldraw[black] (8,0) circle (1.5pt);
\node at (0.2,-0.4) { \footnotesize $\ell_L$}; 
\node at (1.7,-0.38) { \footnotesize $\overline{\ell_L}$}; 
\node at (2.3,-0.4) { \footnotesize $\nu_R$}; 
\node at (3.7,-0.38) { \footnotesize $\overline{\nu_R^{\,c}}$}; 
\node at (4.3,-0.4) { \footnotesize $\ell_L^{\,c}$}; 
\node at (5.7,-0.38) { \footnotesize $\overline{\ell_L^{\,c}}$}; 
\node at (6.3,-0.4) { \footnotesize $\nu_R^{\,c}$}; 
\node at (7.7,-0.38) { \footnotesize $\overline{\nu_R}$}; 
\node at (8.3,-0.4) { \footnotesize $\ell_L$}; 
\node at (9.7,-0.38) { \footnotesize $\overline{\ell_L}$}; 
\node at (3.0,-0.55) { \footnotesize $S_\alpha^\times$}; 
\node at (7.0,-0.55) { \footnotesize $S_\beta^\times$};
\node at (1.9,0.5) { \small $\phi$};
\node at (3.9,0.5) { \small $\phi^T$};
\node at (6.25,0.5) { \small $\phi^*$};
\node at (8.25,0.5) { \small $\phi^\dagger$};
\end{tikzpicture}  
\caption{Two-loop self-energy diagrams for the light $\n_L$ neutrinos giving rise to a 
lepton-antilepton asymmetry in curved spacetime.}
\label{Fig b.3}
\end{figure}

At two loops, we find the three self-energy diagrams illustrated in Fig.~\ref{Fig b.3}.
The corresponding self-energies are:
\begin{align}
\S_{ij}^{(1)}(x,y) ~=~ \int d^4 z \sqrt{-g} &\int d^4 z' \sqrt{-g}\, 
\sum_{\a,\b,k} \left(\l_{i\a} \l_{\a k}^\dagger \l_{k\b} \l_{\b j}^\dagger \right)  \nonumber \\
& \times ~G(x,y)\, G(z,z')\, S_\a(x,z) \,\D(z,z') \, S_\b(z',y) \ , 
\label{b6}
\end{align}
from the `nested' diagram with two $S$ propagators, and 
\begin{align}
\S_{ij}^{(2)}(x,y) ~=~ \int d^4 z \sqrt{-g} &\int d^4 z' \sqrt{-g}\, 
\sum_{\a,\b,k} \left( \l_{i\a} \l_{\a k}^T \l_{k\b}^* \l_{\b j}^\dagger \right)  \nonumber \\
& \times ~G(x,y) \, G(z,z') \, S_\a^\times(x,z) \,\D(z,z') \, S_\b^\times(z',y) \ , 
\label{b7}
\end{align}
and
\begin{align}
\S_{ij}^{(3)}(x,y) ~=~ \int d^4 z \sqrt{-g} &\int d^4 z' \sqrt{-g}\, 
\sum_{\a,\b,k} \left( \l_{i\a} \l_{\a k}^T \l_{k\b}^* \l_{\b j}^\dagger \right)  \nonumber \\
& \times ~G(x,z') \, G(z,y) \, S_\a^\times(x,z) \,\D(z,z') \, S_\b^\times(z',y) \ , 
\label{b8}
\end{align}
for the `nested' and `overlapping' diagrams with two $S^\times$ propagators.
Note that there is no overlapping-type diagram with two $S$ propagators.

Tracing over the light flavours as before, we find the following structure of the
Yukawa couplings for the three diagrams:
\begin{equation}
\tr \left(\S_{ij}^{(1)} - \S_{ij}^{(1)c}\right) ~=~
2i \sum_{\a,\b} \,\Im \left[(\l^\dagger \l)_{\b\a} \, (\l^\dagger \l)_{\a\b} \right]\, I_{\a\b}^{(1)} ~=~ 0 \ ,
\label{b9}
\end{equation}
whereas
\begin{align}
\tr \left(\S_{ij}^{(2)} - \S_{ij}^{(2)c}\right) ~&=~
2i \sum_{\a,\b} \,\Im \left[(\l^\dagger \l)_{\b\a} \, (\l^T \l^*)_{\a\b} \right]\, I_{\a\b}^{(2)} 
\nonumber \\
&=~2i \sum_{\a,\b} \,\Im \left[(\l^\dagger \l)_{\b\a} \, (\l^\dagger \l)_{\b\a} \right]\, I_{\a\b}^{(2)}  \ ,
\label{b10}
\end{align}
and
\begin{equation}
\tr \left(\S_{ij}^{(3)} - \S_{ij}^{(3)c}\right) ~=~
2i \sum_{\a,\b} \,\Im \left[(\l^\dagger \l)_{\b\a} \, (\l^\dagger \l)_{\b\a} \right]\, I_{\a\b}^{(3)}  \ .
\label{b11}
\end{equation}
have the same dependence on the $\l_{i\a}$. Note that 
$\Im \left[(\l^\dagger \l)_{\b\a} (\l^\dagger \l)_{\b\a}\right]$ 
is {\it antisymmetric} in $\a,\b$, so only the antisymmetric part of the dynamical factors 
$I_{[\a\b]}^{(2)}$ and $I_{[\a\b]}^{(3)}$ contributes to the lepton-antilepton asymmetry.

Now, we can prove in general that CPT invariance\footnote{In QFT in curved spacetime, 
we regard the discrete symmetries C, P and T as defined with respect to the local
Minkowski spacetime that exists, due to its pseudo-Riemannian nature, at each point in
spacetime. (See, for example, ref.~\cite{McDonald:2014yfg} for a detailed discussion.)
This is also true of the quantum fields themselves, as evident in the
definition of spinor fields as representations of the SL(2,C) symmetry of the local
Minkowski spacetime.

With this understanding, CPT symmetry and its standard consequences in
QFT for the properties of particles and antiparticles, the 
spin-statistics theorem, {\it etc.} holds independently of the curvature of the background
spacetime, which is {\it not} required to be P or T symmetric.

This may be contrasted with an interesting recent proposal for a ``CPT symmetric'' 
cosmology \cite{Boyle:2018tzc},
in which the gravitational background itself -- an extension to $t<0$ of FRW spacetime --
is taken to be time-reversal symmetric about a single privileged point, interpreted as a 
bounce or creation event. In our interpretation, this ``CPT'' would be viewed as an
{\it environmental} symmetry of the background fields (in this case gravitational), rather
than the fundamental CPT symmetry of the QFT itself. Similar remarks apply to the use of
``CPT'' in the `spontaneous' and gravitational baryogenesis 
papers in refs.~\cite{Davoudiasl:2004gf, Cohen:1987vi}}
and Poincar\'e invariance together imply that the propagation of particles and antiparticles
is identical \cite{McDonald:2015iwt}. 
To see how this is realised in this context, note that in flat spacetime
translation invariance implies that the propagators are functions only of the difference
in coordinates, {\it i.e.} $\D(x,y) \rightarrow \D(x-y)$, {\it etc.} But given this,
we can readily show that the factors $I_{\a\b}^{(2)}$ and $I_{\a\b}^{(3)}$ are symmetric
in $\a,\b$. For example, for diagram (2), translation invariance implies
\begin{align}
I_{\a\b}^{(2)}(x,y) ~&=~\int d^4 z \int d^4 z' \, G(x-y)\,G(z-z')\,S_\a^\times(x-z)\,\D(z-z')
\,S_\b^\times(z'-y)  \nonumber \\
&=~\int d^4 u \int d^4 u' \, G(x-y)\,G(u-u')\,S_\b^\times(x-u)\,\D(u-u')
\,S_\a^\times(u'-y)  \nonumber \\
&=~ I_{\b\a}^{(2)}(x,y) \ ,
\label{b12}
\end{align}
under the change of dummy variables $u = x+y-z'$ and $u'=x+y-z$. 

This is no longer necessarily true in curved spacetime, and indeed we find by explicit
calculation that, provided the Majorana masses of the sterile neutrinos are non-degenerate,
the diagrams with the charge-violating propagators $S_\a^\times$ 
have non-vanishing antisymmetric factors $I_{[\a\b]}^{(2)}$ and $I_{[\a\b]}^{(3)}$.
We therefore find a difference in the self-energies of the light leptons and antileptons
given by 
\begin{equation}
\tr \left(\S_{ij} - \S_{ij}^c \right) ~=~ 2i \sum_{\a,\b} \Im \,[(\l^\dagger \l)_{\b\a} 
(\l^\dagger \l)_{\b\a}]\,
\left(I_{[\a\b]}^{(2)} + I_{[\a\b]}^{(3)}\right) \ .
\label{b13}
\end{equation}
This is the fundamental observation allowing the generation of a lepton-antilepton asymmetry
in curved spacetime through two-loop contributions to the light neutrino propagators 
in this CP-violating BSM model.

\vskip0.2cm

The next step is to determine the coefficients of the effective Lagrangian (\ref{a3})
by explicit evaluation of these diagrams in curved spacetime. Since we only need $L_{eff}$
in the weak curvature regime $\mathfrak{R}/M_\a^2 \ll 1$, the quickest way is to expand the metric
around Minkowski spacetime as $g_{\m\n} = \eta_{\m\n} + h_{\m\n}$ in (\ref{a3})
and evaluate the corrections to the self-energies to $O(h)$. We then match the coefficients
of the corresponding expansions with explicit calculations of the Feynman diagrams above
in flat spacetime but with a single graviton insertion in the propagators or vertices.
Especially at two-loops, these calculations are highly non-trivial, so here we simply refer
to the original papers \cite{McDonald:2014yfg, McDonald:2015ooa, 
McDonald:2015iwt,McDonald:2016ehm, McDonald:2020ghc}
and quote the results:
\begin{equation}
\bigl(a_{ij}, ~c_{ij},~d_{ij} \bigr) ~=~ \frac{1}{12} \frac{1}{(4\pi)^2} \,
\sum_\a \l_{\a j}^\dagger \l_{i\a} \, \frac{1}{M_\a^2} \,
\Bigl(-\frac{4}{3},~\frac{3}{4},\,-3\Bigr) \, \ .
\label{b14}
\end{equation}
and
\begin{equation}
b_{ij} ~=~ \frac{1}{9} \frac{1}{(4\pi)^4} \,\sum_{\a,\b,k} \,\l_{\b j}^\dagger 
\l_{i\a} \l_{\b k}^\dagger \l_{k\a}\,
\frac{1}{M_\a M_\b} \, I_{[\a\b]} \ ,
\label{b15}
\end{equation}
where
\begin{equation}
I_{[\a\b]} ~\sim~ \left(\frac{M_\b}{M_\a}\right)^{2p}\, \log\left(\frac{M_\b}{M_\a}\right) \ ,
\label{b16}
\end{equation}
up to an O(1) numerical factor.

The question of whether the hierarchy parameter $p$ is 0 or 1 was left unresolved 
in ref.~\cite{McDonald:2015iwt}. The most natural result, which holds in all the diagrams
we calculated to a conclusion, is $p=0$. This would accord with the expected decoupling 
of heavy mass intermediate states in the Feynman diagram.
Nevertheless, in calculating the arbitary-momentum, two-loop triangle diagram involved -- which 
appears to be at the limit of known techniques -- we found contributions with $p=1$ which would
require a remarkable cancellation if they were to be absent in $I_{[\a\b]}$. For this reason, we
retain the possibility that $p$ could be 1 in what follows since, as we see in the cosmological
scenarios, the presence of a large sterile mass hierachy dependence could significantly enhance
the ultimate prediction for the baryon-to-photon ratio in this BSM model.

\section{Gravity-Extended Boltzmann Equations in FRW Spacetime}\label{sect 3}

At this point, we have established that loop effects in the BSM model in a gravitational field
may be encoded in the effective Lagrangian (\ref{a3}), with dynamically generated couplings
$a_{ij}, \ldots, d_{ij}$. The next step is to use this effective action to describe the 
gravitational generalisation of the Boltzmann equations which determine the evolution of the
lepton number density in a finite-temperature, expanding universe.

Using the standard Noether procedure with the effective action, we find that the
lepton number current $J^\m = \sum_i \overline{\n_L^{\,i}} \c^\m \n_L^{\,i}$ for the light neutrinos
satisfies the broken conservation equation,
\begin{align}
(1+2cR) D_\m J^\m \,+\, 2a R_{\m\n} D^\m J^\n \,&+\, \Bigl(a + 2c + 
\frac{1}{2}d\Bigr) \partial_\m R J^\m 
\nonumber \\
&+\, 2d\left( \overline{\n_L} D^2 \c.D \n_L
\,+\, \overline{\n_L} \c.\overleftarrow{D} \overleftarrow{D^2} \n_L \right)  ~~\sim ~0 \ ,
\label{c1}
\end{align}
with the same suppression of flavour indices $i,j$ in the couplings, currents and bilinears
employed in (\ref{a3}).  
Since $D_\m J^\m \sim O(\l^2)$, the pre-factor $2cR$ of the $D_\m J^\m$ term must be omitted 
to consistent perturbative order so, up to terms vanishing by the equation of motion, we find
simply,\footnote{From this point, we disregard any light flavour dependence for simplicity 
and take (\ref{c2}) 
as an equation with all the $J^\m$ being flavour singlet currents and with coefficients
$a \sim \tr\, a_{ij}$, {\it etc.} depending on the Yukawa couplings through
$(\l^\dagger \l)_{\a\a}$. } \footnote{See also ref.~\cite{Antunes:2019phe} for related ideas.} 
\begin{align}
D_\m J^\m \,+\, 2a\,R_{\m\n} D^\m J^\n \,+\, 2 \hat{b} \, \partial_\m R J^\m ~\sim~ 0 \ ,
\label{c2}
\end{align}
defining $\hat{b} = \tfrac{1}{2} a + c + \tfrac{1}{4}d$. Note that (\ref{c2}) involves
only the CP even terms -- the CP odd term in the effective action with coupling $b_{ij}$ does 
not contribute to the current conservation equation.

For applications in cosmology, we need to evaluate this in a FRW spacetime. We consider
the spatially flat FRW metric,
\begin{equation}
ds^2\,=\, dt^2 - a(t)^2 \,\d_{ij}  dx^i dx^j \ ,
\label{c3}
\end{equation}
with non-vanishing Christoffel symbols $\Gamma^0_{ij} \,=\, - H \d_{ij}$,
$\Gamma^i_{j0} \,=\, H \d^i{}_j$ and Ricci tensor components
$R_{00} \,=\, 4\pi G \r (1 + 3w)$, $R_{0i} \,=\, 0$, 
$R_{ij} \,=\, -4\pi G\r (1-w) \d_{ij}$. The Ricci scalar is
$R \,=\, -8\pi G\r (1-3w)$ and 
$\dot{R} \,=\, 8\pi G \r\,3H(1-3w)(1+w)$.  
Here, $\r$ is the energy density and the equation of state is $p=w\r$. 
The Hubble constant $H = \dot{a}/a$ is related to $\r$ through the 
Friedmann equation $3H^2 = \r/M_p^2$, where $M_p$ is the reduced Planck mass,
$8\pi G = 1/M_p^2$. 

Evaluating (\ref{c2}) in this background, and identifying the current component $J^0$
as lepton number density $n_L = n_\n - n_{\n^{c}}$, gives the following equation for the
time evolution of $n_L$,
\begin{equation}
(1 + 2a R_{00}) \frac{dn_L}{dt} \,+\, 3H n_L \,+\, 2a R^i{}_i H n_L \,+\, 
2 \hat{b}\,\dot{R}\, n_L \,=\, 0 \ .
\label{c4}
\end{equation}
Writing this to consistent $O(\l^2)$ perturbative order, we find the form
quoted in (\ref{a5}) and substituting for the curvatures gives finally,
\begin{equation}
\frac{dn_L}{dt} \,+\, 3 H n_L ~=~ 3H\frac{\r}{M_p^2} (1+w) 
\bigl[2a - 2 \hat{b} (1-3w)\bigr]\, n_L   \ .
\label{c5}
\end{equation}

To develop this into the full Boltzmann equation, we also need to take account of the
lepton number violating interactions. This is where the non-vanishing equilibrium
density $n_L^{eq}$ induced by the CP odd term in the effective Lagrangian enters.
The $\D L=2$ reactions are shown in Fig.~\ref{Fig b.1}, and the model also allows
contributions from the `inverse decays' $\n_L H \rightarrow \n_R$
at finite temperature. (We neglect $\D L=1$ scattering reactions involving other SM particles 
here for simplicity \cite{Buchmuller:2004nz}.)  
Collecting these effects into a single rate factor $\Gamma = \Gamma_{ID} + 2\Gamma_{\D L=2}$, a 
standard kinetic theory analysis shows that (\ref{c5}) should be
extended to \cite{McDonald:2020ghc}
\begin{equation}
\frac{dn_L}{dt} \,+\, 3 H n_L ~=~ 3H\frac{\r}{M_p^2} (1+w) 
\bigl[2a - 2 \hat{b} (1-3w)\bigr]\, n_L   ~+~ \Gamma \left(n_L - n_L^{eq}\right)\ ,
\label{c6}
\end{equation}
with $n_L^{eq} = \tfrac{1}{3} b\,\dot{R}\, T^2$.

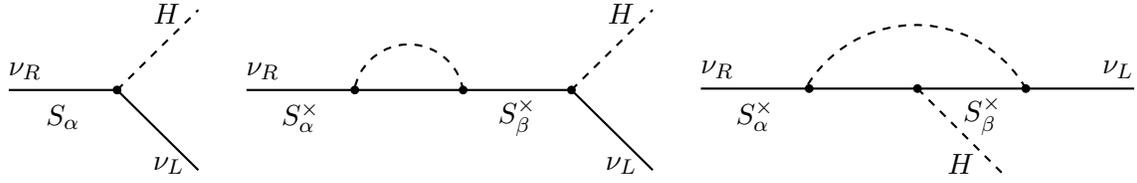
\begin{figure}[h!]
\begin{center}
\begin{tikzpicture}[scale=0.72]
\draw[thick] (0,0) -- (2,0);
\draw[thick,dashed] (2,0) -- (3.5,1.48);
\draw[thick] (2,0) -- (3.5,-1.48);
\filldraw[black] (2,0) circle (2pt);

\node at (0.3,0.35) { \small $\nu_R$}; 
\node at (2.92,1.4) { \small $H$};
\node at (2.95,-1.4) { \small $\nu_L$};
\node at (1.0,-0.5) { \small $S_\alpha$};
\end{tikzpicture}  
\hskip0.35cm
\begin{tikzpicture}[scale=0.72]
\draw[thick] (0,0) -- (2,0);
\draw[thick] (2,0) -- (4,0);
\draw[thick] (4,0) -- (6,0);
\draw[thick,dashed] (6,0) -- (7.5,1.48);
\draw[thick] (6,0) -- (7.5,-1.48);
\draw[dashed,thick] (4,0) arc (10:170:1.02); 
\filldraw[black] (2,0) circle (2pt);
\filldraw[black] (4,0) circle (2pt);
\filldraw[black] (6,0) circle (2pt);
\node at (0.3,0.35) { \small $\nu_R$}; 
\node at (6.92,1.4) { \small $H$};
\node at (6.92,-1.4) { \small $\nu_L$};
\node at (1.0,-0.5) { \small $S_\alpha^\times$};
\node at (5.0,-0.5){ \small $S_\beta^\times$};
\end{tikzpicture}  
\hskip0.35cm
\begin{tikzpicture}[scale=0.72]
\draw[thick] (0,0) -- (2,0);
\draw[thick] (2,0) -- (4,0);
\draw[thick] (4,0) -- (6,0);
\draw[thick,dashed] (4,0) -- (5.6,-1.6);
\draw[thick] (6,0) -- (8,0);
\draw[dashed,thick] (6,0) arc (30:150:2.35); 
\filldraw[black] (2,0) circle (2pt);
\filldraw[black] (4,0) circle (2pt);
\filldraw[black] (6,0) circle (2pt);
\node at (0.3,0.35) { \small $\nu_R$}; 
\node at (4.8,-1.4) { \small $H$};
\node at (7.7,0.35) { \small $\nu_L$};
\node at (1.0,-0.5) { \small $S_\alpha^\times$};
\node at (5.2,-0.5){ \small $S_\beta^\times$};
\end{tikzpicture}  
\end{center}
 \caption{Diagrams for the decay $\n_R^{\,\a} \rightarrow \n_L^{\, i} \, H$ which contribute to
the decay rate asymmetry factor $\varepsilon_\a$.  At $O(\l^4)$ the relevant contribution to the 
$\Gamma(\n_R\rightarrow\n_L\,H)$ decay rate arises from the interference of the tree and
one-loop diagrams shown. 
Similar diagrams, where the $S_\a$ and $S_\a^\times$ type $\n_R$ propagators are switched 
compared to the figure,  give the decay rate for $\n_R^{\,\a} \rightarrow \n_L^{\, i\, c}\, H$.}
\label{Fig c.1}
\end{figure}

Finally, this model -- which after all is a conventional BSM theory for {\it thermal
leptogenesis} -- also includes the standard mechanism for generating a lepton asymmetry
through the out-of-equilibrium decays of the heavy sterile neutrinos,
$\n_R^{\,\a} \rightarrow \n_L^{\,i} H$ and $\n_R^{\,\a} \rightarrow \n_L^{\,i c} H$,
shown in Fig.~\ref{Fig c.1}.
These display an asymmetry dependent on the CP-violating combination of Yukawa 
couplings \cite{Buchmuller:2003gz},
\begin{equation}
\varepsilon_\a \,\simeq\, \frac{3}{16\pi} \sum_{\b \neq \a} 
\frac{{\rm Im}((\l^\dagger \l)_{\a\b}^2)}{(\l^\dagger \l)_{\a\a}} \, \frac{M_\a}{M_\b} \ .
\label{c7}
\end{equation}
arising from the interference of the tree and one-loop diagrams for the decays in 
Fig.~\ref{Fig c.1}. Note that the charge-violating propagators $S_\a^\times$ are crucial in 
establishing the lepton-antilepton asymmetry in these diagrams.

It is usual in discussing leptogenesis to express the evolution in terms of temperature
rather than time, and as in section \ref{sect 1} we introduce the variable $z = M_1/T$
and express the Boltzmann equation in terms of the lepton-to-photon ratio 
$N_L(z) = n_L/n_\c$.\footnote{Since $n_\c \sim T^3$ and $T\sim 1/a$, we have
\begin{equation*}
\frac{dN_L}{dt} ~=~ \frac{1}{n_\c} \left(\frac{dn_L}{dt} \,+\, 3H n_L\right) \ ,
\end{equation*}
and the change of variables gives $d/dz = (1/Hz) d/dt$.}
So in terms of these variables, the final coupled Boltzmann equations for $N_L$ and $N_{\n_R}$
read,
\begin{align}
\frac{d N_{\n_R}}{d z} \,&=\, - D \left(N_{\n_R} \,-\, N_{\n_R}^{eq} \right) \label{c8a} \\
\frac{d N_L}{d z} \,&=\, - D \,\varepsilon_1 \left(N_{\n_R} \,-\, N_{\n_R}^{eq} \right)
\,-\, W \left(N_L \,-\, N_L^{eq}\right)  \, - \, \mathcal{W} \, N_L \ ,
\label{c8b}
\end{align}
where recall that $W(z) = \Gamma/zH$, where $\Gamma(z)$ is the finite-temperature
rate factor appearing in (\ref{c6}) and we similarly define
$D(z) = \Gamma(\n_R \rightarrow \n_L\,H)/zH$. 

The gravitational terms $N_L^{eq}(z)$ and $\mathcal{W}(z)$ in (\ref{c8b}) can be 
read off in terms of the radiatively-generated couplings from (\ref{c6}). We find,
\begin{equation}
\mathcal{W}(z) ~=~ -3 (1+w)\left[2a - 2\hat{b}(1-3w) \right] \,\frac{\r}{M_p^2}\, \frac{1}{z} \ ,
\label{c10}
\end{equation} 
and
\begin{equation}
N_L^{eq}(z) ~=~ (1+w)(1-3w)\, b\, M_1^2\,\frac{H}{n_\c}\, \frac{\r}{M_p^2}\, \frac{1}{z^2} \ .
\label{c9}
\end{equation}

Note here that for a radiation-dominated FRW universe, $w\simeq 1/3$, with the deviation
from this value arising purely from the beta functions characterising
the energy-momentum trace anomaly, $T^\m{}_\m \neq 0$. With standard model fields,
this gives $(1-3w) \simeq 0.1$. This quantum deviation from the classical conformal
symmetry value is essential to realising a non-vanishing $N_L^{eq}(z)$ in (\ref{c9}).
Moreover, in the radiation-dominated case, $H \sim 1/z^2$ while $\r \sim 1/z^4$, 
so both the gravitational terms $N_L^{eq}(z)$ and $\mathcal{W}(z)$ fall sharply as $1/z^5$ 
as the universe cools. 

From (\ref{c10}) we see that whether the radiative curvature corrections encoded in 
$\mathcal{W}(z)$ act to amplify or reduce the lepton number density as the universe evolves 
depends on the sign of the combination $\big[2a - 2 \hat{b} (1-3w)\bigr]$.
Since $(1-3w)$ is small, this essentially depends just on the sign of the coupling $a$; 
negative $a$ tends towards damping (washout) of the lepton number with time, while a 
positive $a$ would imply an amplification. In the BSM model studied here we found in
(\ref{b14}) that $a$ is negative. This was assumed in the picture sketched in
Fig.~\ref{Fig a.1}, where it controls the early, high-temperature evolution in region (a).

\section{Gravitational Leptogenesis in the Early Universe}\label{sect 4}

In this final section, we solve these coupled Boltzmann equations in the BSM model with 
physical neutrino parameters in realistic cosmological settings and discuss the viability
of the gravitational leptogenesis mechanism for generating the observed baryon asymmetry
in the universe today.

\begin{figure}[h!]
\vskip0.5cm
\centering{
\includegraphics[scale=1.2]{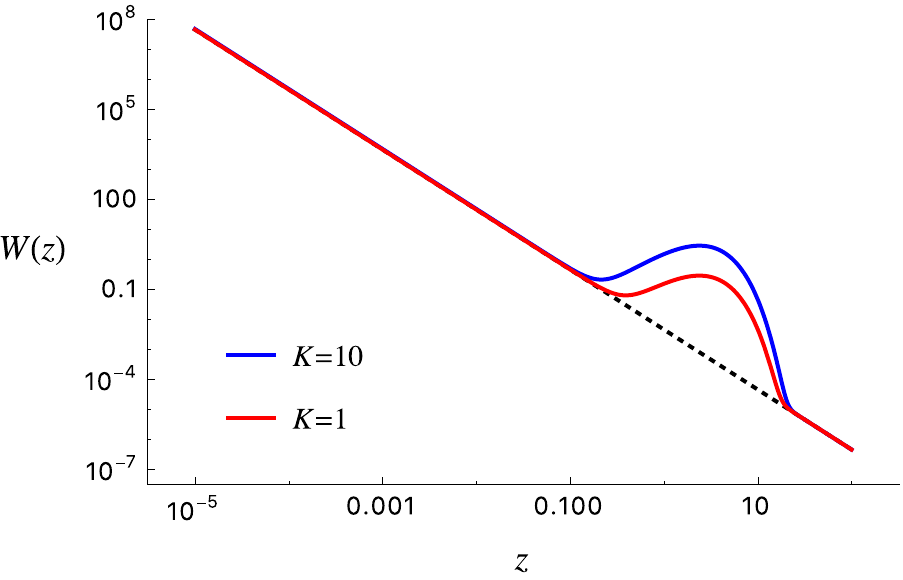}
}
\caption{This shows the dependence on $z = M_1/T$ of the coefficient $W(z)$ in the Boltzmann 
equation for $K=1$ and $K=10$, with $M_1=5\times 10^{10}\,{\rm GeV}$ 
and the neutrino parameters in the text. The resonance peak around $T\simeq M_1$ increases
with $K$.} 
\label{Fig d.1}
\end{figure}

First we need expressions for the conventional finite-temperature rate factors $W(z)$ and
$D(z)$ in (\ref{c8a}), (\ref{c8b}). The necessary results are quoted in 
refs.~\cite{McDonald:2016ehm, McDonald:2020ghc} and are discussed in detail in, for example,
the review \cite{Buchmuller:2004nz}. With standard
definitions\footnote{The definitions 
of the neutrino parameters and notation used
here are explained in refs.\cite{McDonald:2020ghc, Buchmuller:2004nz}. From the see-saw mechanism
we have the sum of the light neutrino masses,
\begin{equation*}
\bar{m}^2 ~=~ v^4 \sum_{\a,\b} \frac{1}{M_\a M_\b} {\rm Re} (\l^\dagger \l)_{\a\b}^2 ~\simeq~
\D m_{sol}^2 + \D m_{atm}^2  \ , 
\end{equation*}
where the `solar' and `atmospheric' masses are $\D m_{sol}^2 = 7.53 \times 10^{-5} \,{\rm eV}^2$
and $\D m_{atm}^2 = 2.44 \times 10^{-3}\,{\rm eV}^2$.
A useful mass scale is set by $m_* = 8\pi (\s/3)^{1/2} \,v^2/M_p = 
1.08 \times 10^{-3}\, {\rm eV}$, where $v = 174\,{\rm GeV}$ is the electroweak scale,  
$M_p = 2.4 \times 10^{18}\,{\rm GeV}$ is the reduced Planck mass, where $8\pi G = 1/M_p^2$,
and $\s = \pi^2 g_*/30$ is the constant appearing in the radiation energy density $\r = \s T^4$,
where $g_*$ is the effective number of relativistic degrees of freedom at temperature $T\sim M_1$.
A key parameter is the combination  
\begin{equation*}
K \,=\, \frac{v^2}{M_1 m_*} \, (\l^\dagger \l)_{11} \ .
\end{equation*}
} 
for parameters related to the light neutrino masses,
we have the following expressions for the lepton number violating rate factor $W(z)$. 
For temperatures well above $T\simeq M_1$, 
\begin{equation}
W(z\ll 1) \, \simeq \, \frac{12}{\pi^2} \, \frac{m_* M_1}{v^2}\, 
\left(\frac{\bar{m}^2}{m_*^2} \,+\, K^2\right) \, \frac{1}{z^2} \ ,
\label{d1}
\end{equation}
with the same formula without the $K^2$ term giving $W(z\gg 1)$.
Near $z\simeq 1$, $W(z)$ shows a resonance enhancement from the intermediate $\n_R^{\,1}$
state, as is evident from Fig.~\ref{Fig b.1}. This is illustrated in
Fig.~\ref{Fig d.1}.
The key features are the asymptotic behaviour $W(z) \sim 1/z^2$, which is a much weaker 
$z$-dependence than the $1/z^5$ of the gravitational terms $\mathcal{W}(z)$ and $N_L^{eq}(z)$,
and the resonance around $z=1$.

Next, the sterile neutrino decay rate factor is given in terms of the Yukawa couplings
and Bessel functions by
\begin{equation}
D(z) \,=\, K\, z \,\frac{K_1(z)}{K_2(z)} \ ,
\label{d2}
\end{equation}
and has an equilibrium value at finite temperature, 
$N_{\n_R}^{eq}(z) = \tfrac{3}{8}\, z^2\, K_2(z)$.

For convenience, we also quote here the formulae for the gravitational terms in a 
radiation-dominated universe with $w\simeq 0.3$ and with the radiatively-induced 
couplings $a, \ldots d$ found in (\ref{b14}), (\ref{b15}).

These are,
\begin{equation}
\mathcal{W}(z) \,\simeq \,  0.075\,  \s^{3/2} \, K \left(\frac{M_1}{M_p}\right)^3\frac{1}{z^5} \ ,
\label{d3}
\end{equation}
and
\begin{equation}
N_L^{eq}(z) \,\simeq \,  0.034\, \s^{3/2} \,\left(\frac{M_1}{M_p}\right)^3\, \sum_{\b\neq1} 
\frac{{\rm Im} (\l^\dagger \l)_{1\b}^2}{(4\pi)^4} \,\left(\frac{M_\b}{M_1}\right)^{2p-1}\, 
\log\left(\frac{M_\b}{M_1}\right) \, \frac{1}{z^5} \ .
\label{d4}
\end{equation}

We have now collected all the ingredients to solve the Boltzmann equations and determine
the evolution of the lepton number asymmetry $N_L(z)$ with temperature in this BSM model,
assuming a radiation-dominated background spacetime.

First, recall that in leptogenesis models, we assume that at much lower temperatures around
the electroweak scale, the lepton number asymmetry generated in the early universe is 
converted through sphaleron processes into a baryon asymmetry. Defining $\eta = n_B/n_\c$
as the baryon-to-photon ratio, the sphaleron conversion in this model gives
$\eta \simeq 0.02\, |N_L|$ \cite{Buchmuller:2002zs}. 
Since the observed value is $\eta \simeq 6 \times 10^{-10}$,
we see that successful leptogenesis requires a final value of $N_L(z)$ for $z \gg 1$ of
$|N_L| \simeq 10^{-8}$. 

\begin{figure}[h!]
\vskip0.3cm
\centering{
\includegraphics[scale=0.78]{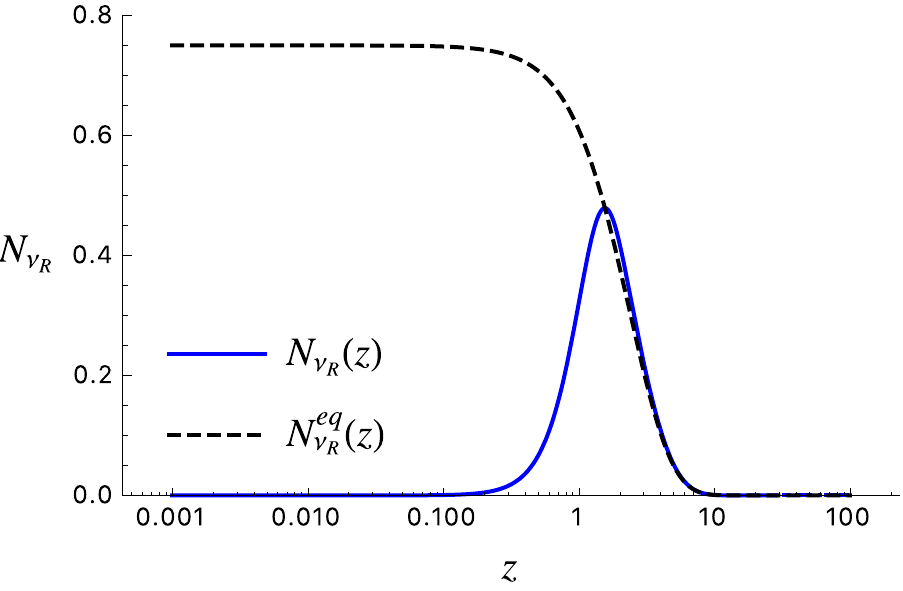}
\hskip0.6cm
\includegraphics[scale=0.58]{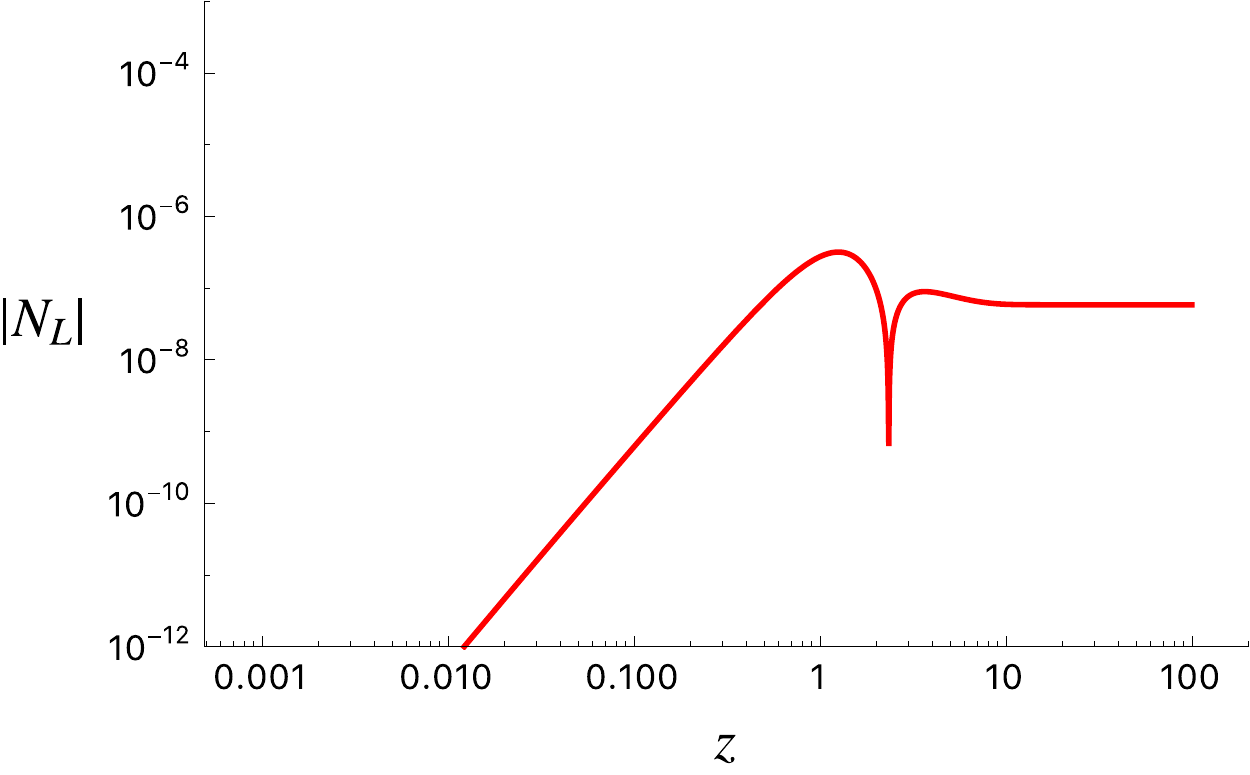}
}
\caption{The left-hand diagram shows the evolution of the sterile neutrino density
starting from an initial condition with $N_{\n_R}(z\ll 1)=0$. 
The right-hand figure shows the corresponding absolute value of the lepton asymmetry induced by 
the out-of-equilibrium decays of $\n_R^{\,1}$.  The cusp in the plot indicates 
that the asymmetry $N_L(z)$ changes sign as the sterile neutrino density $N_{\n_R}(z)$ switches
from under to just over its equilibrium value. 
The parameters here are $M_1 = 5 \times 10^{10}\,{\rm GeV}$ and $K=5$, 
with $\varepsilon_1 = 10^{-6}$ .}
\label{Fig d.2}
\end{figure}
Now consider the Boltzmann equation (\ref{c8a}) for $N_{\n_R}(z)$. Fig.~\ref{Fig d.2} shows
$N_{\n_R}(z)$ rising from an initial condition $N_{\n_R}(z\ll 1) = 0$ at early times, then
slightly overshooting its equilibrium value around $z\simeq 1$ before rapidly being driven back
to equilibrium. Ignoring all other effects in the Boltzmann equation (\ref{c8b}) for $N_L(z)$,
we see that during this out-of-equilibrium phase, the lepton number violating decays
$\n_R \rightarrow \n_L\, H$ generate the lepton asymmetry shown in the right-hand plot
in Fig.~\ref{Fig d.2}. This is of course the familiar mechanism of {\it thermal leptogenesis}
through out-of-equilibrium decays, satisfying the third Sakharov condition. 
Note that the size of the induced asymmetry depends on the decay rate $D(z)$ and, crucially,
on the CP violating parameter $\varepsilon_1$ which is itself sensitive to the phases of
the Yukawa couplings.

\begin{figure}[h!]
\vskip0.5cm
\centering{
\includegraphics[scale=1.2]{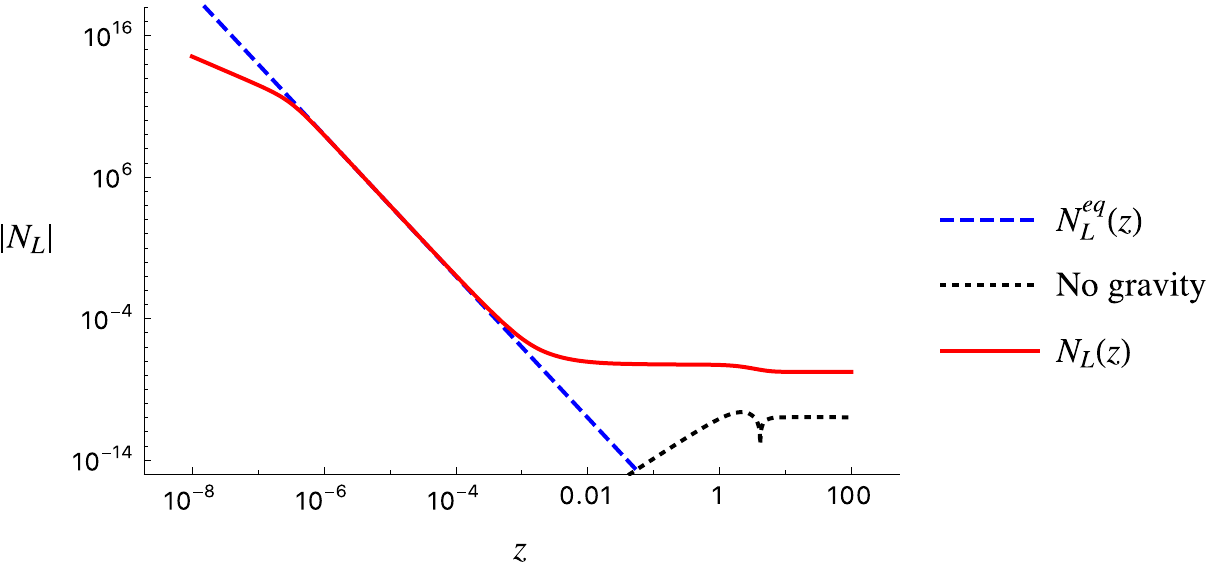}
}
\caption{Dynamical evolution of the lepton number asymmetry $N_L(z)$
in the case of a hierarchy enhancement, $p=1$, with parameters chosen so that the gravitational
leptogenesis mechanism dominates over the out-of-equilibrium $\n_R^{\,1}$ decays.
Here, $M_1=5\times 10^{10}\,{\rm GeV}$, $K=1$, with $M_3= 10^{16}\,{\rm GeV}$,
${\rm Im}(\l^\dagger \l)_{13}^2/(4\pi)^2 = 5\times 10^{-4}$ and $\varepsilon_1 = 10^{-10}$. }
\label{Fig d.3}
\end{figure}

Now consider the evolution of the lepton number asymmetry including the gravitational 
effects encoded in the full Boltzmann equation (\ref{c8b}). This is shown, in the case
of a hierarchy enhancement $p=1$, in Fig.~\ref{Fig d.3}.
From an initial condition $N_L(z\ll 1) \simeq 0$ at ultra-high temperatures,
$N_L$ is initially driven towards its equilibrium value $N_L^{eq}(z) \sim 1/z^5$ by the 
rate factor $W(z)\sim 1/z^2$ for the $\D L \neq 0$ reactions. However, at these early times,
the gravitational `washout' factor $\mathcal{W}(z) \sim 1/z^5$ dominates over $W(z)$
and delays the approach to equilibrium (labelled as region (a) in Fig.~\ref{Fig a.1}).
$N_L$ then follows its equilibrium trajectory (region (b)) until at some lower
temperature ($z\simeq 10^{-3}$ with the parameters shown), the $\D L\neq 0$ reaction rate
falls below the Hubble expansion rate, {\it i.e.}~$W(z) \simeq 1$, and can no longer
maintain $N_L$ in equilibrium. It then decouples and is frozen out at a constant value
(region (c)) as the universe continues to expand and cool. Much later, when $W(z)$ grows
around the resonance at $z\simeq 1$, it is strong enough to temporarily force $N_L$ 
back towards the equilibrium value, which is falling away extremely rapidly, resulting 
in a late dip (region (d)) before settling to its final asymptotic value 
$|N_L| \simeq 10^{-8}$.

The ultimate value of the residual asymmetry $|N_L|$ for temperatures below the Majorana
mass scale, $z \gg 1$, is therefore the result of a competition between the gravitational
leptogenesis mechanism and thermal leptogenesis. Fig.~\ref{Fig d.3} shows the case where
$\varepsilon_1$ is small enough and the gravitational leptogenesis mechanism dominates.

\begin{figure}[h!]
\vskip0.5cm
\centering{
\includegraphics[scale=1.2]{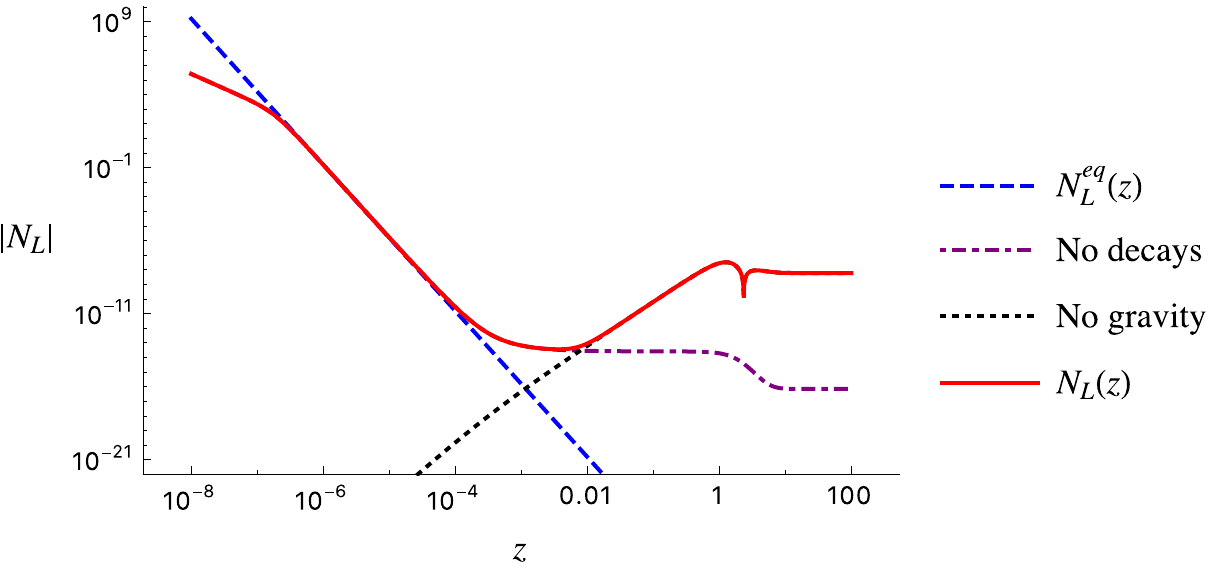}
}
\caption{Dynamical evolution of the lepton number asymmetry $N_L(z)$
in the case of no hierarchy enhancement, $p=0$. Here, the final lepton asymmetry is determined
by the out-of-equilibrium $\n_R^{\,1}$ decays, with the gravitational mechanism producing a steep
rise in the asymmetry at earlier times for temperatures $z \lesssim 0.01$.
Here, $M_1=10^{10}\,{\rm GeV}$, $K=5$, with $M_2= 10^{12}\,{\rm GeV}$,
${\rm Im}(\l^\dagger \l)_{12}^2/(4\pi)^2 = 5 \times  10^{-4}$ and $\varepsilon_1 = 10^{-7}$.}
\label{Fig d.4}
\end{figure}
In contrast, Fig.~\ref{Fig d.4}, where we have assumed no hierarchy enhancement, $p=0$,
and chosen a larger CP violating factor $\varepsilon_1$, shows a scenario in which the
thermal leptogenesis mechanism determines the final value of the asymmetry $|N_L|$. 
Even in this case, however, the gravitational effects describe the early-time evolution
of $N_L(z)$ and predict a large lepton number asymmetry in the early universe. 

\vskip0.2cm
We therefore see from these quantitative examples that radiatively-induced gravitational
leptogenesis provides a mechanism which can generate the observed baryon-to-photon
ratio in the universe today.

\vskip0.2cm
As we have shown, while the RIGL mechanism always determines the evolution of the lepton
asymmetry in the early universe above $T \simeq M_1$, whether the gravitational or sterile 
neutrino decay mechanisms determine the final value of $|N_L|$ depends on the masses and
Yukawa couplings in the BSM model. A comprehensive investigation of the parameter space of 
this model, incorporating flavour effects and varying the sterile neutrino mass spectrum, 
has been given in an interesting recent paper \cite{Samanta:2020tcl}.
In particular, it is shown that exploiting flavour effects, the RIGL mechanism is naturally
able to achieve successful leptogenesis for ranges of the sterile neutrino masses which
are not compatible with standard thermal leptogenesis. This paper also analyses the 
implications of the model parameters favouring gravitational leptogenesis for low-energy 
neutrino phenomenology. 

\begin{figure}[h!]
\vskip0.5cm
\centering{
\includegraphics[scale=1.2]{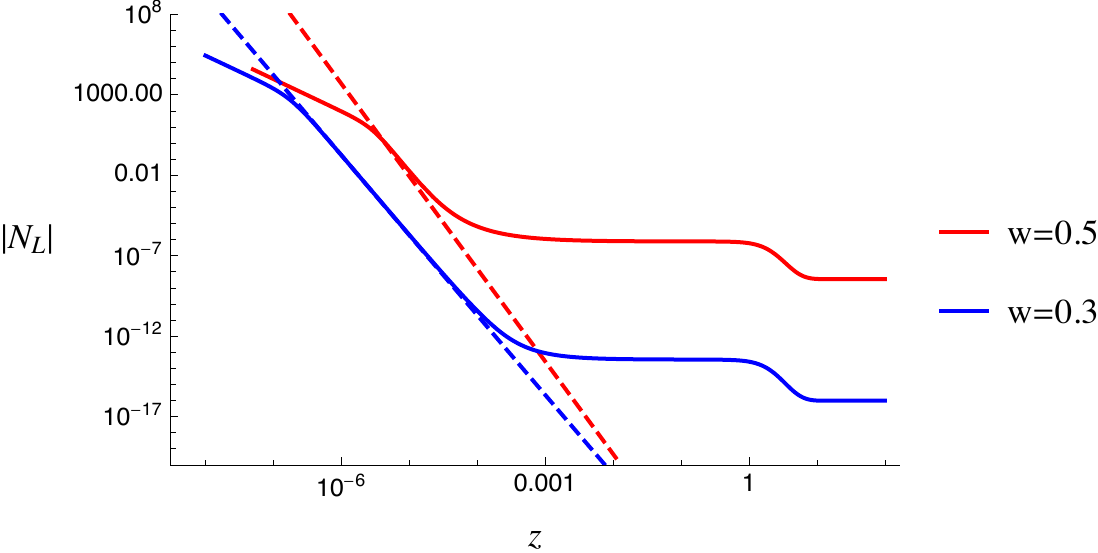}
}
\caption{Dynamical evolution of the gravitationally-induced lepton number asymmetry $N_L(z)$ 
for effective equation of state parameters $w=0.3$ and $0.5$, with $z_* = 0.1$. 
The corresponding equilibrium values $N_L^{eq}$ are shown as the dashed lines.
Decoupling from $N_L^{eq}$ occurs earlier with higher values of $w$, resulting in a larger 
final asymmetry.
The parameters here are $M_1=10^{10}\,{\rm GeV}$, $K=5$, with $M_2= 10^{12}\,{\rm GeV}$,
${\rm Im}(\l^\dagger \l)^2_{12}/(4\pi)^2 = 5 \times  10^{-4}$  
and no hierarchy enhancement, $p=0$.}
\label{Fig d.5}
\end{figure}

Finally, we describe briefly the implementation of the RIGL mechanism in an alternative
cosmological scenario \cite{Davoudiasl:2004gf, McDonald:2016ehm,McDonald:2020ghc}. 
This picture is especially economical in the sense of minimising
the introduction of new fields and interactions, relying on gravitational effects to drive 
the dynamical evolution of the early universe. We therefore suppose that in place of the
usual post-inflationary reheating phase, the relativistic particles giving rise to the
entropy of the universe arise through gravitational particle creation at the transition
from the de Sitter inflationary vacuum to the FRW vacuum at the end of 
inflation \cite{Ford:1986sy, Peebles:1998qn}.
This removes the need for an unknown direct coupling of the inflaton to the standard model
fields to induce reheating. Instead, immediately after inflation the energy density is
dominated by the inflaton component with equation of state $w > 1/3$, which gradually
dilutes relative to the thermalised relativistic particles as the universe expands,
eventually leading, beyond a crossover temperature $z_*$, 
to a conventional radiation-dominated FRW universe.  The stiff
equation of state $w>1/3$ is achieved if the potential at the end of inflation is 
sufficiently steep \cite{Turner:1983he}, the limiting value $w=1$ in which the kinetic energy
dominates being known as `kination'. If the temperature during this
phase is greater than the lightest sterile neutrino mass, $T>M_1$, then gravitational 
leptogenesis occurs during this era.

It is relatively straightforward to trace the dependence on $w$ through the various terms 
in the Boltzmann equations and investigate their solutions. In particuar, the power
dependence of the gravitational terms becomes $\mathcal{W}(z)\sim 1/z^{5-(1-3w)}$
while $N_L^{eq}(z)\sim 1/z^{5 - \tfrac{3}{2}(1-3w)}$, so for higher $w$ the fall-off of
$N_L^{eq}$ is steeper, but decoupling occurs earlier giving a higher final value for $|N_L|$.  
In Fig.~\ref{Fig d.5} we show the dynamical evolution of $N_L(z)$ in an illustrative scenario 
with no hierarchy enhancement but with $w=0.5$, which gives the required asymptotic asymmetry 
$|N_L| \simeq 10^{-8}$, compared to the radiation-dominated case $w\simeq 0.3$ with the
same parameters. 

\vskip0.2cm
Clearly there are many avenues still to explore in developing RIGL as a potential explanation
of the matter-antimatter asymmetry of the universe, both in its implementation in a
variety of BSM models and in different cosmological scenarios. More generally, this work 
once again illustrates the importance of loop effects in QFT in curved spacetime in determining
the physics of the early universe.

\vskip1cm

\noindent{\bf Acknowledgements}
\vskip0.3cm

I am grateful to Jamie McDonald for the collaboration on which this review is based.
This work is supported in part by the STFC theoretical particle physics grant ST/P0005X/1.



\newpage


\begin{thebibliography}{9} 


\bibitem{McDonald:2014yfg}
  J.~I.~McDonald and G.~M.~Shore,
 JHEP {\bf 1502} (2015) 076,
  [arXiv:1411.3669 [hep-th]].

\bibitem{McDonald:2015ooa}
  J.~I.~McDonald and G.~M.~Shore,
 Phys.\ Lett.\ B {\bf 751} (2015) 469,
  [arXiv:1508.04119 [hep-ph]].

\bibitem{McDonald:2015iwt}
  J.~I.~McDonald and G.~M.~Shore,
 JHEP {\bf 1604} (2016) 030, 
  [arXiv:1512.02238 [hep-ph]].

\bibitem{McDonald:2016ehm}
  J.~I.~McDonald and G.~M.~Shore,
 Phys.\ Lett.\ B {\bf 766} (2017) 162,
  [arXiv:1604.08213 [hep-ph]].

\bibitem{McDonald:2020ghc}
  J.~I.~McDonald and G.~M.~Shore,
 JHEP \textbf{10} (2020), 025,
  [arXiv:2006.09425 [hep-ph]].

\bibitem{Davoudiasl:2004gf}
 H.~Davoudiasl, R.~Kitano, G.~D.~Kribs, H.~Murayama and P.~J.~Steinhardt,
 Phys. Rev. Lett. \textbf{93} (2004) 201301, 
 [arXiv:hep-ph/0403019 [hep-ph]].

\bibitem{Khlebnikov:1988sr}
 S.~Khlebnikov and M.~Shaposhnikov,
 Nucl. Phys. B \textbf{308} (1988) 885.

\bibitem{Sakharov:1967dj}
 A.~Sakharov,
 Sov. Phys. Usp. \textbf{34} (1991) no.5, 392.

\bibitem{Fukugita:1986hr}
 M.~Fukugita and T.~Yanagida,
 Phys.\ Lett.\ B {\bf 174} (1986) 45.

\bibitem{Buchmuller:2004nz}
 W.~Buchmuller, P.~Di Bari and M.~Plumacher,
 Annals Phys. \textbf{315} (2005) 305,
 [arXiv:hep-ph/0401240 [hep-ph]].

\bibitem{Cohen:1987vi}
 A.~G.~Cohen and D.~B.~Kaplan,
 Phys. Lett. B \textbf{199} (1987), 251.

\bibitem{Ohkuwa:1980jx}
 Y.~Ohkuwa,
 Prog. Theor. Phys. \textbf{65} (1981) 1058.

\bibitem{Boyle:2018tzc}
 L.~Boyle, K.~Finn and N.~Turok,
 Phys. Rev. Lett. \textbf{121} (2018) 251301,
 [arXiv:1803.08928 [hep-ph]].

\bibitem{Antunes:2019phe}
 V.~Antunes, I.~Bediaga and M.~Novello,
 JCAP \textbf{10} (2019) 076,
 [arXiv:1909.03034 [gr-qc]]

\bibitem{Buchmuller:2003gz}
 W.~Buchmuller, P.~Di Bari and M.~Plumacher,
 Nucl. Phys. B \textbf{665} (2003) 445,
 [arXiv:hep-ph/0302092 [hep-ph]].

\bibitem{Buchmuller:2002zs}
 W.~Buchmuller,
``\textit{Baryo- and leptogenesis (brief summary)},''
 ICTP Lect. Notes Ser. \textbf{14} (2003) 41.

\bibitem{Samanta:2020tcl}
 R.~Samanta and S.~Datta,
 JHEP \textbf{12} (2020), 067,
 [arXiv:2007.11725 [hep-ph]].

\bibitem{Ford:1986sy}
 L.~Ford,
 Phys. Rev. D \textbf{35} (1987) 2955.

\bibitem{Peebles:1998qn}
 P.~Peebles and A.~Vilenkin,
 Phys. Rev. D \textbf{59} (1999) 063505,
 [arXiv:astro-ph/9810509 [astro-ph]].

\bibitem{Turner:1983he}
 M.~S.~Turner,
 Phys. Rev. D \textbf{28} (1983) 1243.


\end{thebibliography}



\end{document}